# Growth, Poverty Trap and Escape

Indrani Bose

Department of Physics, Bose Institute, Kolkata-700009, India

E-mail: indrani@jcbose.ac.in

**Abstract**. The well-known Solow growth model is the workhorse model of the theory of economic growth, which studies capital accumulation in a model economy as a function of time with capital stock, labour and technology-based production as the basic ingredients. The capital is assumed to be in the form of manufacturing equipment and materials. Two important parameters of the model are: the saving fraction $s$ of the output of a production function and the technology efficiency parameter $A$, appearing in the production function. The saved fraction of the output is fully invested in the generation of new capital and the rest is consumed. The capital stock also depreciates as a function of time due to the wearing out of old capital and the increase in the size of the labour population. We propose a stochastic Solow growth model assuming the saving fraction to be a sigmoidal function of the per capita capital $k_p$. We derive analytically the steady state probability distribution $P(k_p)$ and demonstrate the existence of a poverty trap, of central concern in development economics. In a parameter regime, $P(k_p)$ is bimodal with the twin peaks corresponding to states of poverty and well-being respectively. The associated potential landscape has two valleys with fluctuation-driven transitions between them. The mean exit times from the valleys are computed and one finds that the escape from a poverty trap is more favourable at higher values of $A$. We identify a critical value of $A_c$ below (above) which the state of poverty (well-being) dominates and propose two early signatures of the regime shift occurring at $A_c$. The economic model, with conceptual foundation in nonlinear dynamics and statistical mechanics, shares universal features with dynamical models from diverse disciplines like ecology and cell biology.

**Keywords**: Solow growth model, bistability, bimodal probability distribution, potential landscape, poverty trap, mean exit time, randomized control trial

## 1. Introduction

Dynamical systems are characterised by temporal changes in one or more state variables. The changes may be in the size of a population, the amount of capital in an economy, the number of proteins participating in specific cellular events or the concentration of pollutants in the atmosphere. A possible outcome of the dynamics in the long-time limit is that the system reaches a steady state, the fixed-point of the dynamics, in which the temporal rates of change of the state variables fall to zero [1]. In the presence of positive feedback, a sufficiently nonlinear dynamics may result in multistability, i.e., more than one stable steady state. In the case of bistability, two stable steady states are separated by an unstable steady state [1-3]. In terms of a potential landscape defined in state space, the physical picture is that of two valleys separated by a hill. The minima of the valleys represent the stable steady states and the valleys constitute their basins of attraction. The top of the hill, on the other hand, corresponds to an unstable steady state. At a bifurcation point, the parameters governing the dynamics acquire special values with the dynamical system changing from one steady state regime to a different one, say, from bistability to monostability.

The paradigm of bistability (alternative stable states) is of universal origin with examples ranging from ecology, climate science, cell biology to economics and the social sciences [4-9]. In recent years, both field and laboratory experiments have validated the conceptual basis of bistability in diverse dynamical systems [10-12]. In analysing the experimental observations, the issues of special interest are: perturbation-induced transitions from one stable steady state to the other and quantitative early signatures of regime shifts occurring at the bifurcation points [13-14]. In the case of deterministic dynamics, the perturbation could be in terms of sufficiently large added inputs to a relevant state variable so that a transfer between the valleys is possible. In general, the dynamics of most natural processes have a stochastic character with fluctuations (noise) playing a key role in the crossing of the barrier from one valley to the other. Such noise-induced transitions, in certain contexts, have desirable consequences like the generation of population heterogeneity as a bet-hedging strategy or the escape from a valley ("trap") to better conditions in the other valley [15-17].

In this paper, we study a stochastic version of the Solow model [18-20] (Appendix A), a generic model for economic growth. The relevant dynamical variable of the minimal model is the per capita physical capital $k_p$ which increases through production. The capital in the model is defined in terms of accessories like manufacturing equipment and materials. A fraction of the per capita output of production is saved and fully invested in the generation of new capital and the rest is consumed. The negative contribution to the growth rate of $k_p$ comes from two sources: through the wearing out (depreciation) of the accessories and the increase of the population (labour force) at a uniform rate (Appendix A). The dynamics of the original Solow model give rise to one single steady state in which the rates of accumulation and depreciation of the per capita capital balance each other. The addition of new features to the model gives rise to bistable dynamics in an appropriate parameter regime. One such feature is to assume a saving fraction (saving rate) which is not a constant (equation A7) but dpends on the per capita capital $k_p$. In section 2, we introduce the modified Solow model with a $k_p$-dependent saving rate and describe the deterministic dynamics of the model. In the region of bistability, the two valleys represent a poverty trap and a state of economic well-being respectively. In section 3, the stochastic Solow model is analysed in terms of the steady state probability distributions of the per capita capital. The parameter region in which the distribution is bimodal, i.e., a poverty trap exists is identified and the mean exit times from the valleys due to stochastic fluctuations computed. In the last section, the strategies for poverty alleviation, based on the model results, are discussed and the universal features that the economic growth model shares with other dynamical models pointed out.

## 2. Deterministic Solow model and poverty trap

The key determinants of economic growth include capital, labour as well as technology. To take into account the last feature, the output function in (A4) is generalised to $y = Af(k_p)$, where the parameter $A$, termed the technology efficiency (TE) parameter, is a measure of the productivity of the available technology and $k_p$ is the per capita capital (income). A fraction $s$ of the output is saved and invested for the generation of new capital with the per capita investment $i = sy = sAk_p^\alpha$. The major focus of Solow's economic growth model was on economies with a single steady state, an outcome of a constant saving fraction $s$ and a concave production function of the CD form (see Appendix A). Solow, additionally, pointed out that one could obtain multiple steady states with an appropriately chosen nonconcave form for the production function. With the inclusion of the TE parameter $A$, the time evolution of $k_p$ is given by the rate equation

$$.\frac{dk_p}{dt} = A\,s\,k_p^\alpha - (an + \delta)k_p \quad (1)$$

where $an$ is the population (labour) growth rate and $\delta$ the depreciation rate of $k_p$. Figure 1(a) (left panel) illustrates the case of a single steady state for the parameter values $A = 10.5, s = 0.2, \alpha = 0.4, an = 0.5, \delta = 0.5$. The figure shows the plots of the rate of increase in the per capita capital $k_p$, termed "saving" (the first term in the r.h.s. of equation (1)) and the rate of net depreciation, termed "depreciation" (the second term in the r.h.s. of equation (1)) versus $k_p$. The intersections of the two curves represent the fixed points of the dynamics with a trivial fixed point at $k_p^* = 0$ (unstable steady state) and a second fixed point at $k_p^* = 3.431$ (stable steady state). The existence of multistability, specifically, bistability, is of significant interest as it could signify the existence of a poverty trap separated from a state of economic well-being by an unstable steady state. A purely concave curve representing the "saving" term (per capita investment) can intersect the linear curve representing "depreciation" at most at two points, only one of which corresponds to a stable steady state. With a nonconcave "$S$-shaped" functional form of the "saving" term, bistability, i.e., the coexistence of a poverty trap with a state of economic well-being is possible. In the following, we will discuss two possible origins of the S-shaped form: one based on a sigmoidal form of the production function and the other on a sigmoidal form of the saving rate as a function of $k_p$. Note that both the production function and the saving rate appear in the per capita investment ("saving") term and the assumption of a sigmoidal form in any one of the two factors fulfills the condition for the occurrence of bistability in a specific parameter regime. A typical sigmoidal curve is convex-concave, convex below the point of inflection and concave above and provides a natural basis for an $S$-shaped growth dynamics.

In this paper, we assume that the the saving fraction, assumed to be constant in the original Solow model, has a sigmoidal dependence on the per capita capital $k_p$. The existing economic literature [21] provides considerable evidence of a positive relationship between the saving fraction (rate) $s$ and the per capita capital (income) $k_p$ with an increase in $k_p$ resulting in a higher value of $s$. In the closed economy of the Solow model, the income generated has two components, saving and consumption. It is the saving component which provides the capital for investment so that a higher rate of saving results in a higher rate of capital accumulation. For low values of $k_p$, the value of $s$ is expected to be small as the available resources are mostly utilised in consumption. The saving rate rises as the per capita income increases but at a decreasing rate till an asymptotic value is reached. The saving fraction thus has a low value for low $k_p$ values, increases as $k_p$ increases and then saturates at high levels. This type of functional variation, characteristic of a sigmoidal function, has been shown to lead to bistability in the Solow model [22], giving rise to a poverty trap (the "savings trap" model). Assuming the saving fraction to be a function of $k_p$, the rate equation (1) becomes

$$\frac{dk_p}{dt} = A\, s(k_p)\, k_p^\alpha - (an + \delta)k_p \qquad (2)$$

where the production output is $Ak_p^\alpha$ of which a fraction $s(k_p)$ is saved and invested. The functional dependence of $s(k_p)$ is assumed to be sigmoidal having the form of the four-parameter logistic nonlinear regression [23]:

$$s(k_p) = s1 + \frac{(s2 - s1)k_p^{s3}}{k_p^{s3} + d^{s3}} \qquad (3)$$

The sigmoidal form is analogous to the chemical dose-response curve in the form of a Hill function. The parameter $s1$ ($s2$) represents the saving fraction when $k_p$ is zero (very large), with the parameter $d = k_p$, $s(k_p)$ is halfway between the minimum and maximum saving fractions, $s1$ and $s2$ respectively.

Figure 1(b) (right panel) shows the plots of the "saving" and "depreciation" terms (the first and the second terms, respectively, in the r.h.s. of equation (2)) versus $k_p$. The saving function parameters are $s1 = 0.1, s2 = 0.2, s3 = 10.0, d = 2.0$. The other parameter values are the same as in the case of figure 1(a). For this set of parameter values, the plots in figure 1(b) intersect at four fixed points with the "saving" curve a nonconcave function of $k_p$. Ignoring the trivial fixed point at $k_p^* = 0$, the dynamics represent bistability with two stable steady states $k_{p1}^* = 1.089$ and $k_{p3}^* = 3.431$ separated by an unstable steady state $k_{p2}^* = 1.931$. The two stable steady states represent states of poverty and economic well-being respectively. The region $0 < k_p < k_{p2}^*$ constitutes the basin of attraction of the stable steady state $k_{p1}^*$ so that, for any initial state located in this basin, the steady state $k_{p1}^*$ is attained in the long time limit. Similarly, the region $k_{p2}^* < k_p$ constitutes the basin of attraction of the stable steady state $k_{p3}^*$. The basin of attraction of $k_{p1}^*$ constitutes the poverty trap, as defined in economic literature [24,25]. In the case of deterministic dynamics, escape from the trap is possible through inputs of additional per capita capital $k_p$ so that the threshold value $k_{p2}^*$ is crossed. The supply of sufficient additional input capital in the form of aid (" big push") is one of the best-known strategies suggested so far for poverty alleviation [26,27]. The bistability arises due to the presence of a positive feedback: increased (decreased) savings increase (decrease) investment and thus more (less) physical capital generation which in turn increases (decreases) savings and so on. Above the threshold set by the unstable steady state, the self-reinforcing mechanism results in a steady state which is the state of well-being and below the threshold the steady state is that of poverty.

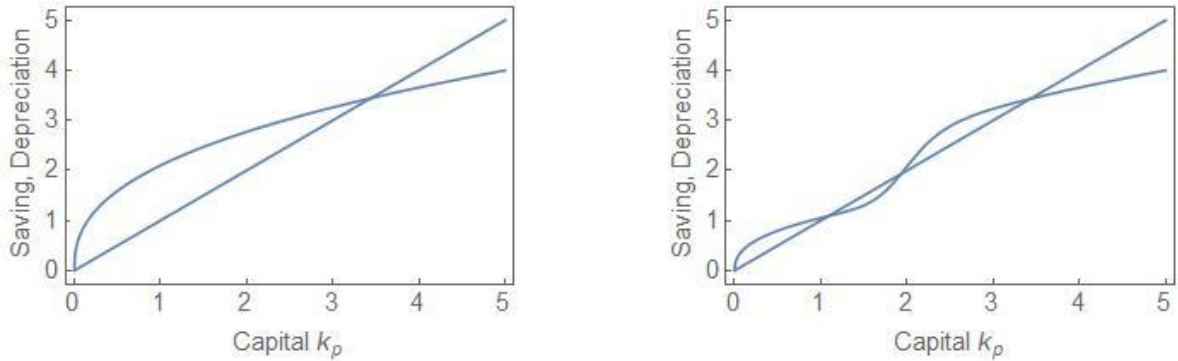

Figure 1. Saving (investment) and depreciation rates versus $k_p$ for (a) $s1 = s2 = 0.2$. (b) $s1 = 0.1, s2 = 0.2$. The other parameter values are $A = 10.5, \alpha = 0.4, s3 = 10.0, d = 2.0, \delta = 0.5, an = 0.5$.

In development economics, the evidence of a poverty trap in a real economy as an outcome of bistable dynamics is a highly debatable issue [28]. A recent pioneering experiment by Balboni et al. on a limited segment of an economy provides concrete evidence of the threshold-based phenomenon of bistability [27]. The experiment involved about 6000 households in Bangladesh with cows constituting the capital (asset). Large-scale randomized asset transfers were undertaken and the data collected for an eleven-year period. From the collected data, a threshold level of initial assets was identified. Families with initial assets below the threshold were mostly engaged in unproductive occupations as domestic servants and agricultural labourers whereas those with initial assets above the threshold could accumulate assets through productive occupations like livestock rearing and land cultivation. The distribution of the productive assets amongst the households in the villages was of a distinct bimodal character. As part of a poverty alleviation strategy, asset transfers were made to a random fraction of households trapped in poverty. The data collected four and eleven years after the transfer clearly established that the beneficiaries for whom the transfer amounts were inadequate for the threshold to be crossed could not get out of the poverty trap, whereas with adequate transfers, household assets of the beneficiaries rose above the threshold and escape from poverty was possible through engagement in productive occupations. After the eleven-year period, the dynamics of the two groups diverged. The results are consistent with the threshold-based growth of an economy and its convergence to one of two possible stable steady states, economic decline (poverty) and economic well-being. The usefulness of a "big push" in asset transfer as a poverty alleviation strategy was supported by the collected data.

The proposal that a bistable dynamics give rise to a poverty trap gets adequate support from a comprehensive analysis of the data collected in the Bangladesh field experiment [26,27]. The data provide information about the transition equation, $K_{t+1} = \Phi(K_t)$, connecting accumulated capital (assets) $K$ of a household at time $t + 1$ to the capital at time $t$. The equation is the same as equation (2) but with time discretized in appropriate intervals. The fixed points in the discrete case are given by the intersection points of the curve described by the transition equation in $K(t + 1) - K(t)$ space, with the $45°$ straight line on which $K(t + 1) = K(t)$. In the continuous time rate equation approach, the fixed poins are the solutions of the equation $\frac{dk_p}{dt} = 0$ (figure 1). As pointed out by Balboni et al. [27], the transition equation, obtained from field data collected at successive time points, can be categorized into three types: Type I with a single stable steady state, Type II with two stable steady states separated by an unstable steady state with the transition curve having the ubiquitous $S$-shape and Type III, with the two stable steady states separated by a discontinuity rather than an unstable steady state. The last case corresponds to a binary form for the saving fraction $s$ as a function of the per capita capital $k_p$, i.e., $s(k_p) = s_L$ for $k_p \leq k_c$ and $s(k_p) = s_H$ for $k_p > k_c$, where $k_c$ sets a threshold, the location for a jump discontinuity [28]. Both Type II and Type III transition equations yield two stable steady states, one of which could define a poverty trap. In the Type III case, capital accumulation occurs just above the jump discontinuity. On the other hand, in the Type II case, capital does not accumulate close to the point respresenting the unstable steady state. The evidence from the Bangladesh data conforms to the Type II transition equation with an $S$-shaped curve, (note the appearance of the analogous $S$-shaped curve, in the continuous time case, in figure 1(b)), describing the relationship between the current and past assets. In this scenario, households with initial assets below (above) the threshold, describing the unstable steady state, would progressively become poorer (richer) till the stable steady state is reached. Since the asset paths below and above the threshold move in opposite directions, the majority of households would fall into two distinct groups, with per capita capital considerably below (poverty trap) and considerably above (state of economic well-being) the threshold set by the unstable steady state. Relatively few households would have their asset stocks close to the threshold. This is exactly what Balboni et al. [27] find from an analysis of their data, thus establishing the veracity of the poverty trap model with Type II transition equation. As pointed out by them, different mechanisms could give rise to the $S$-shaped transition equation. In the case of a Solow-type growth dynamics, the shape could arise from a nonconcave production function or if the saving fraction $s$ is an increasing function of the per capita capital [27]. The functional form of $s(k_p)$, as shown in equation (3), fulfills the criteria for the observation of Type II transition dynamics. A similar functional form of $s(k_p)$ has been proposed in a recent study [22] classifying appropriate strategies for poverty alleviation in multidimensional poverty trap models interlinking Solow-type economic growth with socio-ecological factors.

The other possibility, that of a nonconcave production function mapping per capita capital into per capita output, has been discussed in a number of studies [29, 30] with the specific functional form being given by the sigmoid :

$$f(k) = \frac{\mu k^p}{1 + \gamma k^p} \qquad (4)$$

where $\mu, \gamma$ are nonnegative and $p \geq 2$. The CD production function, used in the Solow model, is recovered with $\mu = 1$ and $\gamma = 0$. While the CD production function is concave throughout the range of $k_p$, the production function in equation (4) is convex (concave) for $k < k_f$ $(k > k_f)$ where $k_f$ denotes the point of inflection of the sigmoid. A constant term, $c, (c > 0)$, may be added to the r.h.s. of equation (4) to take into account the possibility that production is possible without capital $(c = f(0))$, which Solow mentions in his Nobel Prize lecture. In this paper, we study economic growth based on equations (2) and (3), i.e., consider the $S$-shaped dynamics required for bistability to arise from a graded dependence of the saving fraction on $k_p$. This assumption is not restrictive as similar results are obtained if one alternatively uses equation (4) (sigmoidal production function) to obtain bistability.

In development economics literature, a number of mechanisms has been suggested for the generation of single and multiple equilibria in capital growth dynamics [31]. The different possibilities include single equilibria representing states of poverty and economic well-being respectively and multiple equilibria describing the coexistence of the states of poverty and economic well-being in a single economy. The production technology, utilised in capital generation, is a key determinant of the nature of the equilibria attained by the economy in the long run. In the Solow model, the parameter $A$ provides a measure of the efficiency of the production technology. We adopt the "potential landscape" approach for a direct visualization of the possible attractors of the growth dynamics. Figure 2 shows how the potential (stability) landscape, defined by the potential function, $U$ versus $k_p$, is transformed as the TE parameter $A$ is changed. The potential function $U$ is computed from the relation

$$\frac{dk_p}{dt} = F(k_p) = A\, s(k_p)\, k_p^\alpha - (an + \delta)k_p = -\frac{dU(k_p)}{dk_p} \qquad (5)$$

The successive values of $A$ from left to right in the top and bottom rows are $A = 9.09, 9.9, 10.1, 12.12$ respectively.

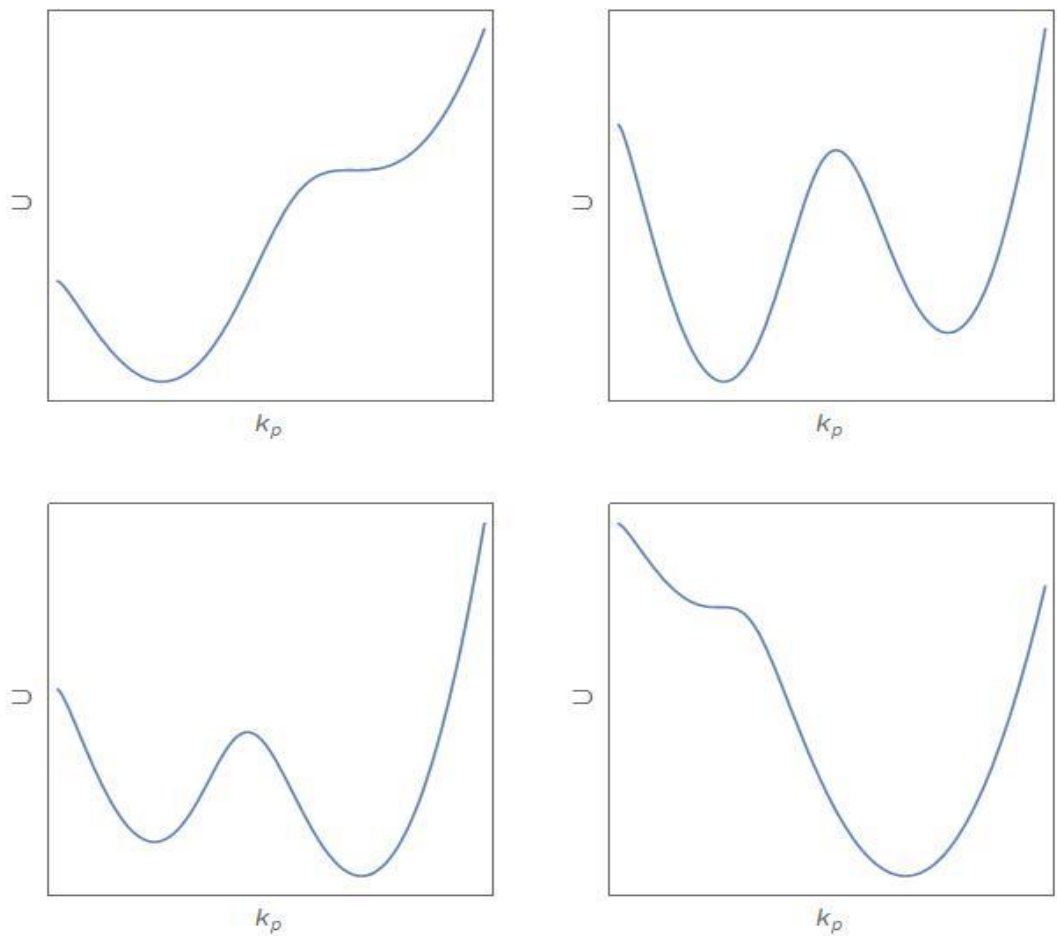

Figure 2. Potential function $U$ versus $k_p$ for $A = 9.09, 9.9$ (top row, left to right) and $A = 10.1, 12.12$ (bottom row, left to right). The other parameter values are: $s1 = 0.1, s2 = 0.2, s3 = 10.0, \alpha = 0.4, d = 2, \delta = 0.5, an = 0.5$.

As the TE parameter $A$ changes from $A = 9.09$ to $A = 12.12$, the state of the economy changes from the single equilibrium state of poverty (monostability), through a region of bistability in which the states of poverty and well-being coexist, to finally a single equilibrium state of well-being ( monostability). Technological efficiency/progress thus appears to be a crucial factor in promoting equilibria respresenting states of economic well-being. For $A = 9.09$, the level of technology is not sufficiently productive to generate non-poor economic states whereas for a sufficiently high level of technology ($A = 12.12$), the poverty trap no longer exists.

Figure 3 depicts the variation of the stability index $\lambda$ of a stable steady state versus $k_p$ as a function of the TE parameter $A$ with $\lambda$ computed using the relation

$$\lambda = \left| \frac{dF(k_p)}{dk_p} \right| (k_p = k_p^*) \right| \qquad (6)$$

The stability diagram has two branches with $A$ in the range of values 9.1 to 12.12. The branches are computed for the two stable steady states $k_{p1}^*$ (state of poverty) and $k_{p3}^*$ (state of well-being) respectively. The poor (non-poor) economic state gradually loses stability as $A$ increases (decreases) from the value 9.1 (12.12) to completely lose stability ($\lambda = 0$) at the bifurcation point close to 12.12 (9.1). The parameter values are the same as in the case of Figure 2.

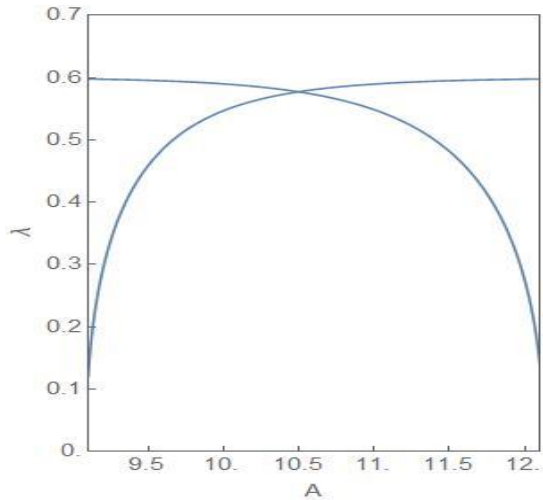

Figure 3. The stability index $\lambda$ versus the TE parameter $A$ with the parameter values the same as in Fig. 2.

While comparing technology versus saving in driving economic growth, one has to take into account the fact that the saving fraction is limited to a restricted range of values due to practical limitations whereas technology efficiency may undergo steady improvements in time with no conceivable upper limit.

**3. Noise-induced escape from poverty trap**

The early economic literature provides some examples of studies on stochastic generalizations of the Solow growth model in both discrete and continuous times. These initial studies contributed significantly in constructing a conceptual and mathematical basis for the existence, uniqueness and stability of the steady state probability distributions of the capital/labour ratio [32, 33]. The sources of uncertainty considered included random variations of the TE parameter $A$ through "technology shocks" and fluctuations in the key parameters like the size of the labour force (population), saving and depreciation rates [33,34]. The mathematical treatments of the stochastic models were in certain cases beset with boundary value problems which led to a gradual loss of interest in the models in the later-day economic literature. The steady state probability distributions of these earlier models were unimodal with a single stable steady state in the deterministic versions of the models. More recent stochastic models have proposed new mechanisms for the generation of bimodal steady state probability distributions with the modes representing states of poverty and well-being [35]. In most of the deterministic models of poverty trap, the condition of bistability is achieved by assuming binary values for either the saving fraction $s$ or the TE parameter $A$ [25, 28], separated by a threshold value. The threshold point sets a point of discontinuity above which economic agents (households, firms) are able to accumulate capital and reach the state of well-being under steady state conditions whereas below the point of discontinuity, the agents end up in the state of poverty [25, 27]. A more realistic framework requires a graded dependence of, say $s$, on $k_p$ which has been shown to result in three steady states, two stable steady states separated by an unstable steady state [22]. The Bangladesh randomized control trial (RCT) data on household asset dynamics [27] share similar features with the models of graded dependence of the saving fraction on $k_p$, namely, the existence of an unstable steady state and an $S$-shaped function representing the asset dynamics ( figure 1(b)).

The stochastic rate equation for the Solow model, studied by us, is obtained by adding a noise term to the deterministic rate equation in equation (5). With an additive noise,

$$\frac{dk_p}{dt} = F(k_p) + \sigma \xi(t) \qquad (7)$$

where $\sigma^2$ is the noise intensity and $\xi(t)$ represents a white noise with zero mean and time correlation given by $<\xi(t)\xi(t')> \sim \delta(t-t')$. The saving fraction $s$ in the model has a sigmoidal dependence on $k_p$ as shown in equation (3). In the equivalent formalism of the Fokker-Planck Equation (FPE), the stochastic rate equation deals directly with the probability distribution $P(k_p, t)$ with $P(k_p, t)dk_p$ interpreted as the probability of finding $k_p$ in the interval $(k_p, k_p + dk_p)$ at time $t$. The FPE is given by [36,37]

$$\frac{\partial}{\partial t}P(k_p, t) = -\partial_{k_p}[F(k_p)P(k_p, t)] + \frac{\sigma^2}{2}\partial_{k_p k_p}P(k_p, t) \qquad (8)$$

From equation (8), the steady state probability distribution is given by

$$P(k_p) = N \, exp\left(\left(\frac{2}{\sigma^2}\right) \int F(u) \, du\right) \qquad (9)$$

where $N$ is the normalisation constant and the integral is evaluated at $k_p$. The integral has an analytic solution given by

$$\int F(u) du = \frac{As1 u^{1+\alpha}}{1+\alpha} - (\delta + an)\frac{u^2}{2} + Ad^{-s3}\frac{(-s1+s2)}{1+s3+\alpha} u^{1+\alpha+s3} H \qquad (10)$$

where $H$ is the hypergeometric function,

$$H = Hypergeometric \, _2F_1\left[1, \frac{1+\alpha+s3}{s3}, 1+\frac{1+\alpha+s3}{s3}, -\left(\frac{u}{d}\right)^{s3}\right] \qquad (11)$$

Figure 4 shows the steady state PDF $P(k_p)$ (equation (9)) versus $k_p$ for $A = 9.9$ and $A = 10.1$ respectively. With a small change in the TE paramete r $A$, the state of the economy changes from a predominance of poverty (economic decline) to a predominance of well-being (economic progress). In analogy with the potential function $U$, defined in equation (5), one can define a stochastic potential $\varphi$ from the steady state probability distribution $P(k_p)$ as [36, 37]

$$P(k_p) = N \, exp\left(-\frac{2}{\sigma^2}\varphi(k_p)\right) \qquad (12)$$

The maxima of the steady state probability distribution $P(k_p)$ correspond to the minima of the stochastic potential $\varphi(k_p)$. In the case of additive noise, the deterministic potential $U(k_p)$ coincides with the stochastic potential $\varphi(k_p)$ modulo an inessential constant. The extrema of the steady state probability distribution thus have an exact correspondence with the valleys and the hilltop of the potential landscape in the deterministic case and can be considered as the stochastic representations of the macroscopic steady states. The unique steady state distribution with peaks centred around the deterministic stable steady states is stable in the sense that the system dynamics converge over time to the same probability distribution independent of the initial distributions. This non-dependence on history is a unique feature of ergodic dynamics investigated in economic literature in connection with general economic growth theories and macroeconomics [35]. The ergodicity is an outcome of stochastic fluctuations which bring about transitions between the basins of attraction so that a dynamical exploration of the full state space is possible. In the absence of fluctuations, i.e., deterministic dynamics, the location of the stable steady state in one valley or the other depends on the initial state (history matters) with no inherent scope of transitions between valleys resulting in nonergodicity. In the stochastic case, if fluctuations are small, the system state, a metastable state, wanders around the local minimum of a specific valley. A single large fluctuation or a sequence of small but synergistic fluctuations brings about transitions across the barrier between the two valleys. The "mean exit time" (MET) from a basin of attraction provides a quantitative measure of the resilience of the attractor. The greater the magnitude of the MET from a valley, the larger is the peak height of the probability distribution, associated with that valley, as the system spends more time in the vicinity of the local minimum of the valley. A difference in the heights of the twin peaks in the steady state probability distribution (figure 4) indicates the difference in the magnitudes of the MET from the two valleys and hence their resilience.

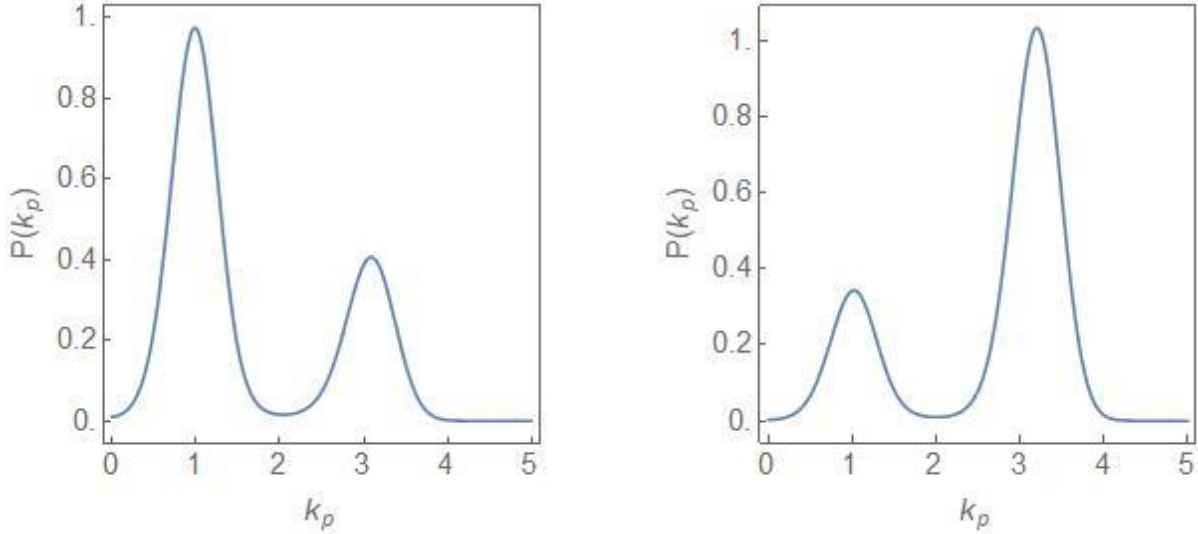

Figure 4. The steady state probability distribution $P(k_p)$ versus $k_p$ for (left) $A = 9.9$ and (right) $A = 10.1$ with $\sigma = 0.3$. The other parameter values are the same as in Figure 2.

A bimodal probability distribution is symptomatic of the observed polarity in the world distribution of per capita capital $k_p$, referred to in economic literature, as "club convergence" or "twin peaks", i.e., the coexistence of multiple equilibria [38, 39]. The MET is defined to be the average time $T$ it takes for a stochastic trajectory to leave a specific valley for the first time with $x_0$ as the initial state. To compute the METs from the two valleys, we use the following equation, derived from the backward FPE [40],

$$D_1(x_0)\frac{dT}{dx_0} + D_2(x_0)\frac{d^2T}{dx_0^2} = -1 \qquad (13)$$

In the case of the Solow model, the system variable $x = k_p$ and the initial state $x_0 = k_{p0}$. The "drift" $D_1(x_0)$ and the "diffusion" $D_2(x_0)$ are given by

$$D_1(x_0) = F(k_p), D_2(x_0) = \frac{1}{2}\sigma^2 \qquad (14)$$

The appropriate boundary conditions for each basin of attraction are: absorbing boundary condition $T(x_0) = 0$ at the border, defined by the unstable fixed point, between the two basins of attraction and reflecting boundary condition $\frac{dT}{dx_0} = 0$ (no change in $T(x_0)$) at the other two boundaries, namely, left boundary of left basin and right boundary of right basin. The absorbing boundary condition takes care of the fact that the MET at the unstable fixed point is zero and the reflecting boundary conditions confine the system to the observed range of values of $x$. Figure 5 shows the plots of the MET $T$ versus the initial state $x_0$ for the same values of the TE parameter $A$ as in the case of figure 4. The values are $A = 9.9$ (left) and $A = 10.1$ (right). For $A = 9.9$, the valley associated with the poverty trap is deeper than the valley describing the states of well-being. The MET from the poverty trap is correspondingly larger. The situation is reversed when the parameter $A = 10.1$.

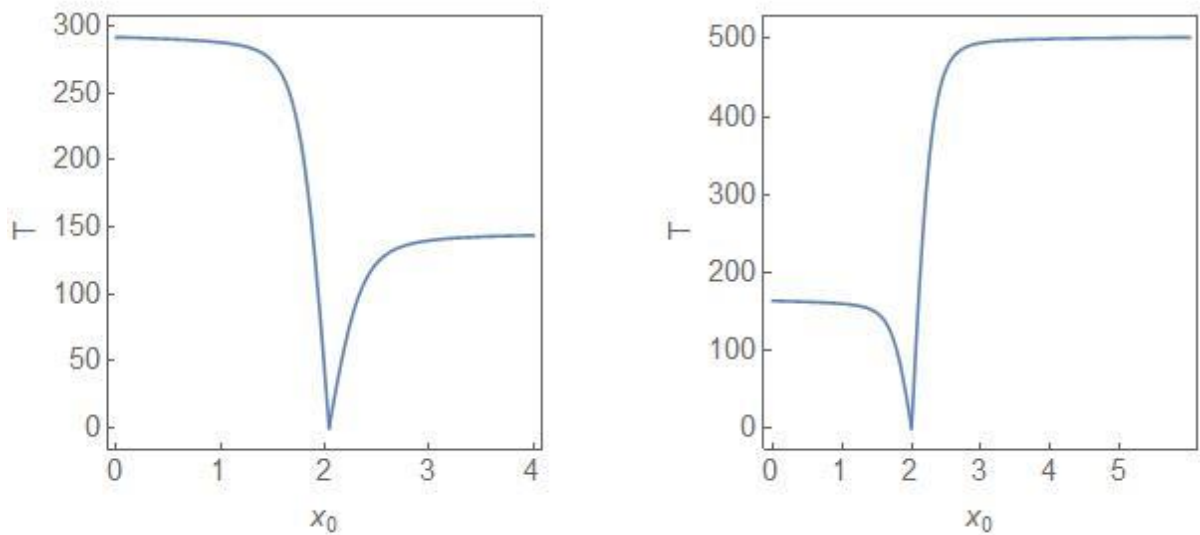

Figure 5. The MET $T$ versus the initial state $x_0$ (the initial per capta capital) for (left) $A = 9.9$ and (right) $A = 10.1$ with $\sigma = 0.3$. The other parameter values are as in figure 2.

As the figure shows, the magnitude of the MET falls sharply in the vicinity of the unstable fixed point becoming zero at the fixed point itself. The stochastic component of the dynamics allows for escape from the poverty trap, the escape is more favourable higher the value of the parameter $A$.

As the value of $A$ is raised, the poverty trap becomes more and more shallow with an associated loss of resilience. The MET from the poverty trap is much lower than that from the valley of well-being. Figure 6 depicts this situation for $A = 10.5$ and $\sigma = 0.3$.

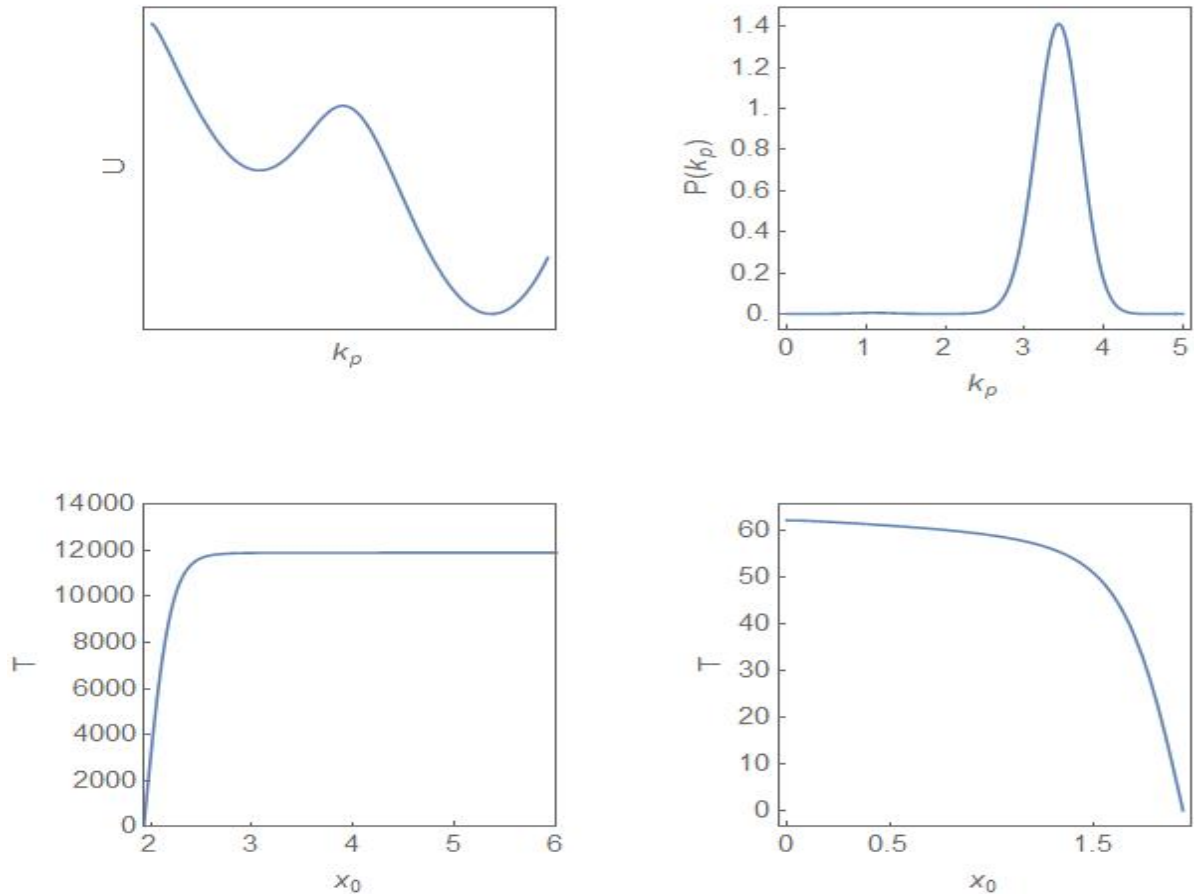

Figure 6. (top row) Potential function $U$ versus $k_p$ and the steady state probability distribution $P(k_p)$ versus $k_p$. (bottom row) MET from the valley of well-being (left) and MET from the poverty trap (right). The parameter values are $A = 10.5$ and $\sigma = 0.3$ with the other parameter values the same as in figure 2.

The peak height of the steady state probability distribution corresponding to the poverty trap is very small and is not resolved in the figure. The MET from the poverty trap is a about 1000-fold smaller in magnitude than that in the case of the other valley. In this case, only a tiny minority of the economic agents is located in the poverty trap. Due to the possibility of fluctuation-driven escapes from the valleys, the fate of an agent may change in the course of time, from the state of poverty to that of well-being and vice versa, albeit at different rates. The probability distribution, on the other hand, attains an invariant form in the steady state,

For $A = 9.9$, the twin peaks in the steady state probability distribution are of equal height so that the associated valleys in the potential landscape are of equal depth (figure 7). The corresponding METs are of similar magnitude. This value of $A = A_c = 9.9$ is of special note as $A_c$ separates two regimes: for $A < A_c$, the economy enters the regime of decline with the predomonance of poverty as $A$ decreases whereas for $A > A_c$, the regime of economic progress prevails as the parameter $A$ increases.

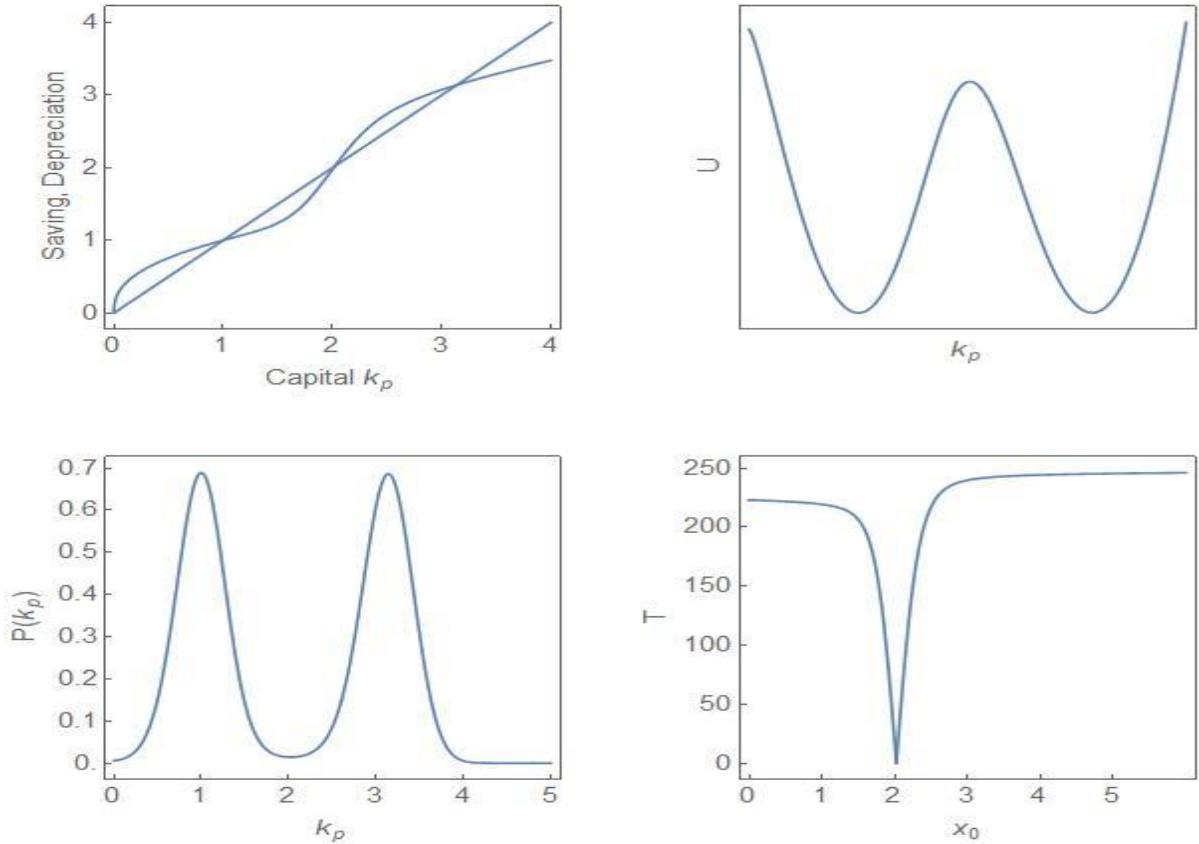

Figure 7. (top row) Saving, depreciation versus $k_p$ (left), potential function $U$ versus $k_p$ (right). (bottom row) Steady state probability distribution versus $k_p$ (left), METs versus initial state $x_0$ (right). The parameter values are $A = 9.99, \sigma = 0.3$ with the other parameters as in figure 2

The special value of $A_c = 9.99$, which is not a bifurcation point, can be thought of as a sort of tipping point between the two steady state regimes of economic progress and decline. One early quantitative signature of the regime shift in this case is the rising variance of the steady state distribution (figure 8) as the tipping point is approached with the variance reaching its maximum value at $A = A_c$. The mean of the steady state distribution as a function of $A$ has the sharpest rise at the tipping point $A_c$.

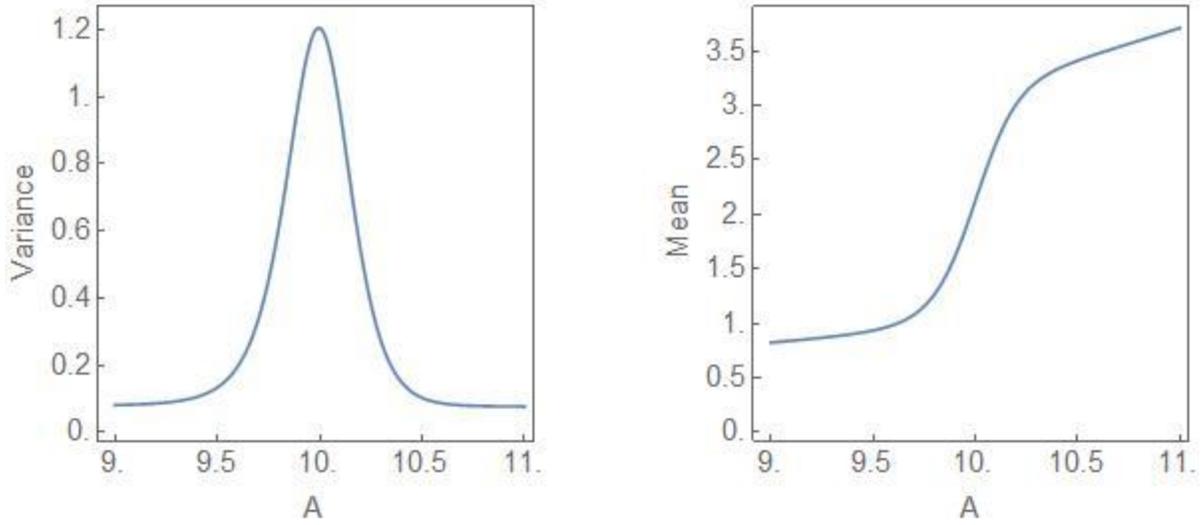

Figure 8. (left) Variance of steady state probability distribution $P(k_p)$ versus $A$. (right) Mean of $P(k_p)$ versus $A$. The parameter values are as in figure 2 and $\sigma = 0.3$.

Figure 9 shows the effect of decreasing and increasing the magnitude of noise $\sigma$ on the steady state probability didtribution $P(k_p)$ versus $k_p$. The reference distributions are those in figure 4 for which $\sigma = 0.3$. The top (bottom) row depicts the distributions for decreased (increased) noise $\sigma = 0.25$ ($\sigma = 0.4$). All other parameter values are as in figure 4.

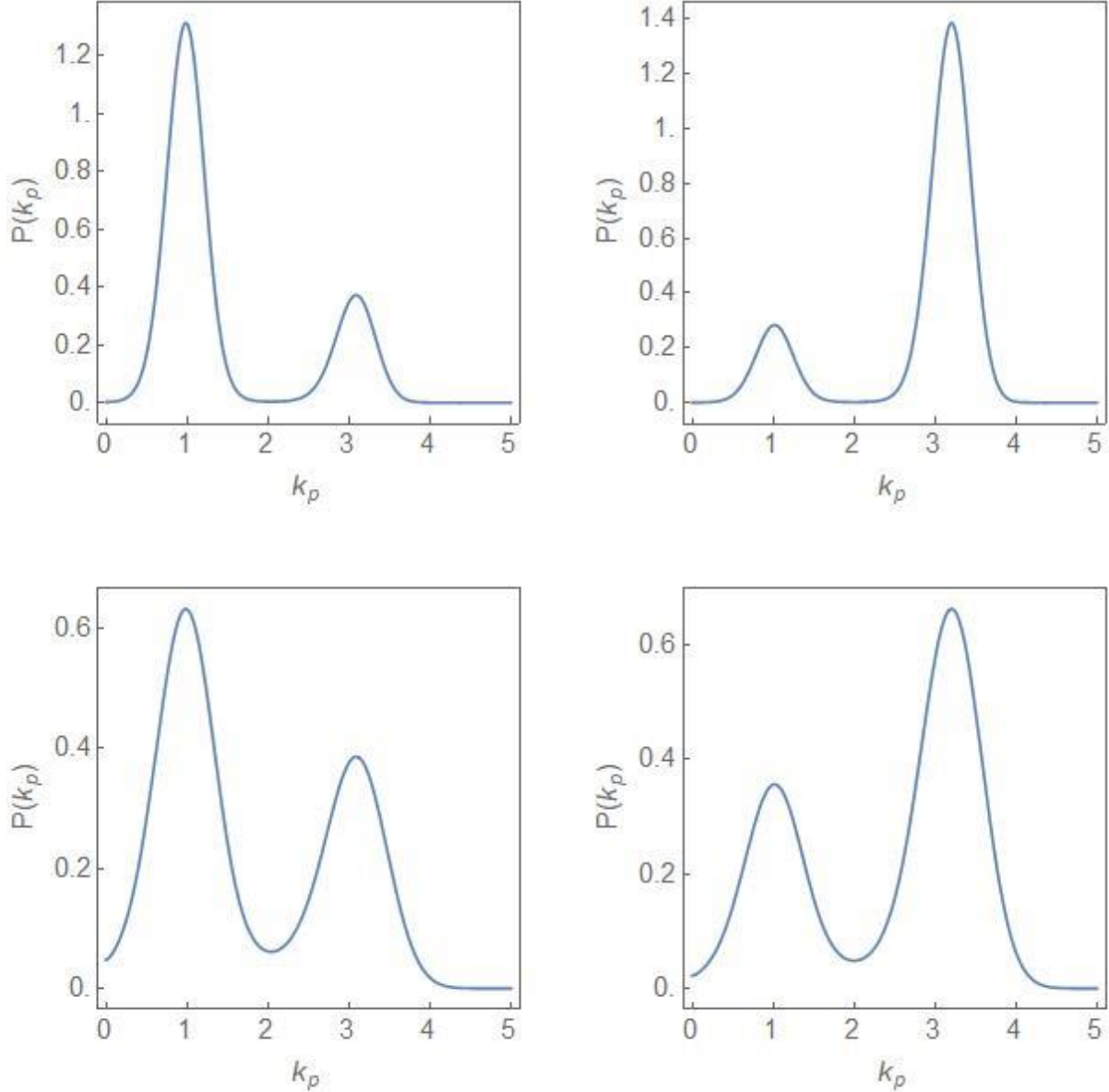

Figure 9. $P(k_p)$ versus $k_p$. The top (bottom) row corresponds to noise magnitude $\sigma = 0.25$ ($\sigma = 0.4$). The left (right) column distributions are for $A = 9.9$ ($A = 10.1$) as in figure 4. All the other parameter values are the same as in figure 4.

A comparison of figures 4 and 9 shows that with decresed noise, $\sigma = 0.25$, fluctuation-induced transitions to the less prominent valley are reduced in number resulting in a decrease and increase of the peak heights of $P(k_p)$ associated, respectively, with the less and more prominent valleys in the potential landscape. On the other hand, with increased noise, $\sigma = 0.4$, $P(k_p)$ has a small but finite value at intermediate levels of $k_p$ and fluctuation-driven transitions to the shallower valley increase in number. Figure 10 shows the MET $T$ versus the initial state $x_0$ diagrams corresponding to the potential landscapes associated with the steady state probability distributions displayed in figure 9. For reduced (increased) noise, $\sigma = 0.25$ ($\sigma = 0.4$), the METs increase (decrease) by a significant amount from those shown in figure 4 for $\sigma = 0.3$.

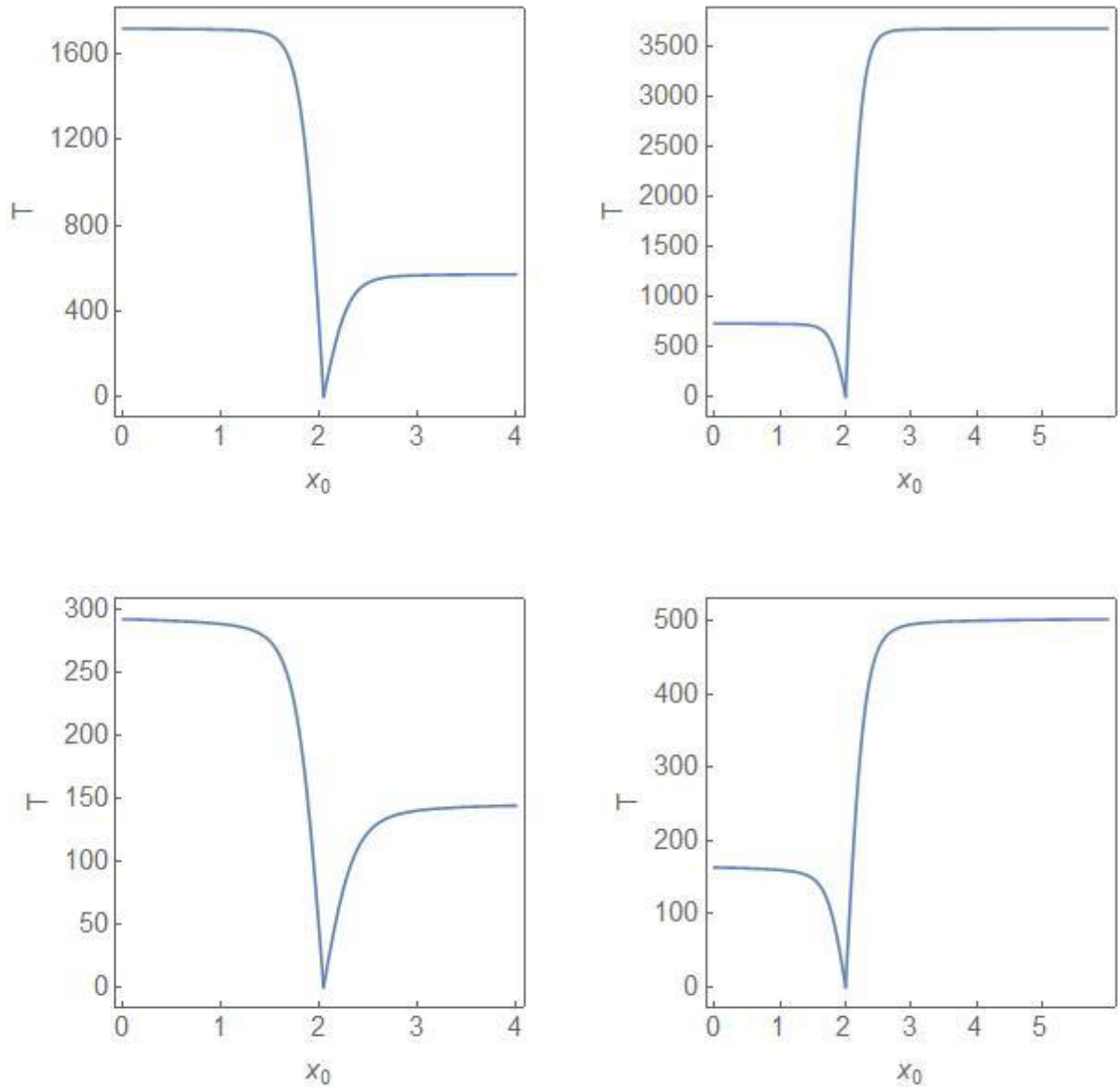

Figure 10. MET $T$ versus $x_0$. The top (bottom) row corresponds to noise magnitude $\sigma = 0.25$ ($\sigma = 0.4$). The left (right) column distributions corresponds to $A = 9.9$ ($A = 10.1$) as in figure 4. All the other parameter values are the same as in figure 4.

A summary of the major results obtained from our study of the stochastic Solow model and their implications is as follows. The Solow model is a minimal model of economic growth which captures some of the universal features of world economies, the most important of which is convergence towards a steady state. In the case of the stochastic Solow model, the convergence is towards an invariant probability distribution. On the inclusion of additional features like per capita capital-dependent saving fraction, the existence of a poverty trap in a parameter regime can be demonstrated (figure 1). The bistable potential landscape (figure 2) shows the coexistence of two valleys corresponding to the states of poverty and well-being. The potential landscape offers a physical picture for escape from a poverty trap to states of well-being via a sufficiently "big push" in terms of additional input capital. The push is needed to cross the barrier separating the poverty trap from the valley of well-being. The transition from the region of bistability to that of monostability occurs at two bifurcation points (figures 2, 3), the bifurcation parameter being the TE parameter $A$. These bifurcation points have been described as "tipping points" in economic [31, 41] as well as interdisciplinary literature [4,37]. As the magnitude of $A$ decreases towards the lower bifurcation point, the valley associated with economic well-being starts losing stability and it becomes increasingly difficult to escape the poverty trap. The reverse situation is true if the magnitude of $A$ increases towards the upper bifurcation point. Thus the magnitude of the push, big or small, depends on the magnitude of $A$. The point to take specially note of is that technological efficiency/productivity plays a key role in the eradication of poverty.

Our stochastic Solow model includes the simplest type of noise, namely, additive noise the magnitude of which does not have a functional dependence on the state variables of the system. The noise could arise from the random processes associated with an economy and could have a sizeable magnitude in unusual situations which gives rise to economic instabilities (war, epidemic, natural disaster, market crash, periods of lockdown during the Covid-19 pandemic etc.). The additive noise provides the simplest representation of the randomness associated with a dynamical system. In a specific parameter regime, the steady state probability distribution of the per capita capital, in the stochastic Solow model, has a bimodal character. A similar bimodal asset distribution characterises the Bangladesh data as well as the data pertaining to a number of other economies [27, 35]. A fluctuation-driven escape from a poverty trap is possible irrespective of the initial per capita capital depending on the barrier height and the intensity of noise. The reverse process of falling into a poverty trap is also possible, though the METs of the two types of transitions could be vastly different (figure 6). In the case of our stochastic Solow model, an analytic solution of the FPE is possible with the steady state probability distribution $P(k_p)$ involving the hypergeometric function (equations (7) - (11)). The computation of the steady state probability distributions and the METs for different values of the bifurcation parameter $A$ and the magnitude of noise $\sigma$ provides quantitative estimates of the propensity of an economy towards states of poverty and well-being and how easy or difficult it is to escape from a poverty trap. A significant result of our study is the identifaction of a critical parameter value $A_c$, which is not a bifurcation point and which separates two distinct economic regimes, one in which poverty dominates and the other in which states of well-being are prevalent. If the TE parameter $A$ is close to $A_c$, the METs from the valleys, depicting the states of poverty and well-being, are of similar magnitude, making a two-way reversal in fortunes a quite common occurrence as in the case of, say, a protracted war. The chance of a fluctuation-driven escape from a poverty trap increases as A increases beyond $A_c$. At the critical value, the variance of the steady state probability distribution, $P(k_p)$, reaches its maximum value and the mean of the distribution has its sharpest increase. These quantities may thus provide precursor signatures of regime shifts from the dominance of the states of poverty to that of the states of well-being and vice versa. In Ref. [41], technological progress has been identified as the critical factor in bringing about a sharp change, from the pre-industrial to the post-industrial age, in the societal and economic living conditions, akin to the drastic changes one observes in a phase transition.

4. **Concluding remarks**

The concept of poverty traps is central to development economics with a poverty trap defined as "any self-reinforcing mechanism which causes poverty to persist" [35]. The modified Solow growth model studied by us highlights the importance of technological productivity/efficiency in the eradication of a poverty trap. The escape from a poverty trap, in the absence of aid, may be driven by stochastic fluctuations and such transitions become more favourable as the trap becomes less deep due to technological progress. The computed MET gives an estimate of the time required for making an escape. With all the other parameters remaining the same, the progress (decline) of an economy occurs if A is above (below) a critical value $A_c$. While a more thorough analysis of the results of the model from the perspective of economic growth is needed, we would like to point out the universal features that the model shares with some of the dynamical models studied in ecology and cell biology.

The common mathematical framework of the interdisciplinary models utilises the fundamental concepts and techniques of nonlinear dynamics and statistical mechanics. The basic ingredients of the models are the same: positive feedback, nonlinear dynamics and stochastic fluctuations. The first two ingredients are essential for bistability and the third promotes heterogeneity in the form of a bimodal probability distribution in the steady state. In the case of ergodic models, different initial probability distributions converge in the course of time to the same invariant probability distribution in the steady state, as has been mentioned in economic literature [35] and experimentally verified in cell biology experiments [42]. In the context of economic growth, a desirable option is monostability with a single valley of well-being. This regime is separated from the regime with a single poverty trap by a regime of bistability/bimodality. In specific cell biological situations, the living cell prefers the option of bimodality so that the cell population is heterogeneous with two prominent subpopulations, the outcome of a bet-hedging strategy (not investing one's total money in a single stock or not putting all one's eggs in the same basket). The heterogeneity in the cell population is achieved through fluctuation-driven transitions between the valleys. Stochasticity in this case has a desirable consequence. The advent of single-cell experimental techniques in cell biology has made it possible to test the conceptual basis of the dynamical models in actual experiments [43]. Field and laboratory scale experiments in ecology serve a similar purpose, linking observations to model predictions [4,11,13,14]. Several experimental studies provide early signatures of regime shifts in diverse dynamical systems at the bifurcation or the so-called tipping points. In our study on the stochastic Solow model, we have demonstrated the existence of two such signatures of an approaching tipping point, $A_c$, which is not a bifurcation point, at which the states of poverty and well-being are evenly balanced but away from which a tilt favouring one or the other regime (economic decline versus economic progress) occurs. The bifurcation point transitions are analogous to equilibrium phase transitions and have been discussed/analyzed utilizing the criticality perspective [41, 44, 45]. In economics and the social sciences, the opportunities for field experiments are limited but cross-disciplinary perspectives could facilitate in identifying the universal features, opening up in the process new pathways for exploration.

## Acknowledgement

IB acknowldges the support of NASI, Allahabad, India, under the Honorary Scientist Scheme.

## Appendix A

The Solow Growth Model

The Solow Growth (SG) model, first proposed in 1956, is a minimal model of economic growth in a closed economy (absence of government or international trade) [18-20]. The model focuses on economic growth through the accumulation of capital which includes accessories like manufacturing equipment, bulldozers, semiconductors etc., to be utilised in the generation of new capital. The model is characterised by a production function (PF), $F(K, L)$, with two categories of inputs, capital $K$ and labour $L$ (population fully employed as labour). The production function has the well-known Cobb-Douglas (CD) form with the output $Y$ given by

$$Y = F(K, L) = K^\alpha L^{1-\alpha} \qquad (A1)$$

The time dependence of the quantities $L, K$ and $Y$ is not explicitly shown but is implied. The PF has the feature of constant returns to scale, i.e.,

$$rY = F(rK, rL) \qquad (A2)$$

with $r$ any positive number. If the capital and labour are both multiplied by the factor $r$, the output $Y$ is also multiplied by the same factor. This feature allows one to write equation (A1) in terms of per capita (per worker) quantities. Defining

$$k = \frac{K}{L}, y = \frac{Y}{L} \qquad (A3)$$

one obtains the reduced equation

$$y = f(k) = k^\alpha \qquad (A4)$$

with $f(k) = F\left(\frac{K}{L}, 1\right)$. The output of the PF constitutes the "income" (goods in the Solow model) in the economy. A fraction $s$ of the per capita output $y$ is saved and fully invested (investment equals saving) in acquiring new equipment and accessories thus raising the capital stock and the rest of the output is consumed. Thus,

$$y = c + i \qquad (A5)$$

where $c = (1-s)y$ is the per capita consumption and $i = sy = sk^\alpha$, is the per capita investment. The investment has a positive contribution to the growth rate, $\frac{dk}{dt}$, of the per capita capital. The negative contributions to the growth rate comes from two sources. The labour force in the Solow model is assumed to grow at a constant rate,

$$\frac{dL}{dt} = nL \qquad (A6)$$

where $n$ is the per capita growth rate of the labour force. The negative contribution in this case is given by the second term on the r.h.s. of the equation, $\frac{1}{k}\frac{dk}{dt} = \frac{1}{K}\frac{dK}{dt} - \frac{1}{L}\frac{dL}{dt}$, $k = \frac{K}{L}$, which from equation (A6), is $-n$. Since the whole population is fully employed as labour, $n$ is also the per capita population growth rate. The second negative contribution, $-\delta k$, is due to capital depreciation through the wearing out of accessories. The time evolution of the per capita capital is thus given by the differential equation

$$\frac{dk}{dt} = sk^\alpha - (n+\delta)k \qquad (A7)$$

This is the core equation of the SG model and leads to the interesting conclusion that the growing economy reaches a steady state $\left(\frac{dk}{dt}\right) = 0$ in the long run irrespective of the initial state of the economy. The steady state values of $k$ and $y$ are given by

$$k^* = \left(\frac{s}{n+\delta}\right)^{\frac{1}{1-\alpha}}, \qquad y^* = (k^*)^\alpha = \left(\frac{s}{n+\delta}\right)^{\frac{\alpha}{1-\alpha}} \qquad (A8)$$

A rise in the saving fraction $s$, and consequently the investment, increases the steady state per capita capital. On the other hand, an increase in the population growth rate and/or the depreciation rate reduces the same. The growth of the economy occurs only during the time interval before the steady state is reached. The growth rate slows down as the per capita capital increases, i.e., the economy becomes more developed. Some of the predictions of the SG model find support in the economic growth data of various countries [18-20]. The SG model, as described here, has been modified in a number of later-day models of economic growth.